**Solitonic Phase in Manganites.**

Luis Brey[1] and P.B.Littlewood[2]

[1]*Instituto de Ciencia de Materiales de Madrid (CSIC), Cantoblanco, 28049 Madrid, Spain.*

[2]*Cavendish Laboratory, Cambridge University, Madingley Road, Cambridge CB3 0HE, UK*

**Whenever a symmetry in the ground state of a system is broken, topological defects will exist. These defects are essential for understanding phase transitions in low dimensional systems[1]. Excitingly in some unique condensed matter systems the defects are also the low energy electric charge excitations. This is the case of skyrmions in quantum Hall ferromagnets[2] and solitons in polymers[3]. Orbital order present in several transitions metal compounds[4-6] could give rise to topological defects. Here we argue that the topological defects in orbital ordered half doped manganites are orbital solitons. Surprisingly, these solitons carry a fractional charge of $\pm e/2$, and whenever extra charge is added to the system an array of solitons is formed and an incommensurate solitonic phase occurs. The striking experimental asymmetry in the phase diagram as electrons or holes are added to half doped manganites[7-12], is explained by the energy difference between positive and negative charged solitons. Contrary to existent models that explain coexistence between phases in manganites as an extrinsic effect[13-14], the presence of inhomogeneities is naturally explained by the existence of solitonic phases. The occurrence and relevance of orbital solitons might be a general phenomena in strongly correlated systems.**

Manganites have the composition $(R_{1-x}A_x)MnO_3$ where R denotes rare earth ions and A is a divalent alkaline ion. In these oxides $x$ coincides with the concentration of holes moving in the $e_g$ orbital bands of the Mn ions that ideally form a cubic structure. The properties of manganites depend on the competition between the kinetic energy tending to delocalise the carriers and localization effects such as the electron phonon interaction and the antiferromagnetic (AF) coupling between the Mn core spins. Manganites with large hopping amplitude between Mn ions have a metallic ground state[15] whereas in manganites for which localization effects are more important the ground state develops charge and orbital order[16-18]. The metallic phases are ferromagnetic whereas ordered phases present AF order in general. Nevertheless, unexpected coexistence of charge order and ferromagnetism has also been observed[19].

The main physic of manganites resides in the large Hund's coupling, which forces the spin of the holes moving the $e_g$ bands towards a local alignment with to the Mn core spin texture[20]. Microscopic models based in this effect and including Jahn-Teller coupling ($\lambda$) and AF interaction between the Mn spins ($J_{AF}$) describe appropriately the *uniform* phases that appear near half doping as well as the temperature induced phase transitions. In general there is a good experimental[7-10] and theoretical[21-24] agreement on the commensurate nature of orbital, charge and spin ordered phases occurring in manganites at half doping, $x=1/2$. In the commensurate phase, the wave vector of the modulation, **q**, is commensurate with **a**\*/2 being **a**\* the reciprocal lattice vector. However, when moving away from half doping the competition between phases produces a strong asymmetry with respect to the addition of electrons or holes. It is experimentally well established that added electrons maintain the system commensurate; however, when doping with holes some experiments indicate phase separation between commensurate phases[11], whereas other experiments state the existence of incommensurate phases, with a modulation given by a wave vector **q**=$(1-x)$**a**\*[12]. This is a fundamental issue in the physics of manganites that still remains to be understood. Combining general topological arguments and realistic microscopic calculations, we will show that the existence of incommensurate phases and the experimental striking asymmetry with respect to the addition of electrons or holes are a consequence of the existence of charged solitons in the orbital order of the half doped manganites. The different phases presented in this work and their energies have been obtained by solving a two-orbital double exchange hamiltonian coupled to Jahn-Teller phonons. For details, see Methods.

In half doped manganites the Jahn-Teller coupling, the directionality in the hopping amplitude and the antiferromagnetic exchange interaction conspire to create an exotic phase of charge, orbital and spin order. Magnetically the Mn ions present AF order of CE type[25]: the $x$-$y$ planes are antiferromagnetically coupled and into the planes, see Fig.1, the magnetic structure consists of ferromagnetic zigzag chains coupled antiferromagnetically. The steps, both in the $x$ and $y$ directions, contain three Mn ions. Due to the large Hund's coupling and to the conservation of the spin in the hopping process, in the CE phase the carriers can only move along the zigzag chain. Therefore the kinetic energy of the whole system is the sum of the kinetic energy of the individual chains. In the $x$-$y$ plane the hopping amplitude between Mn ions depends on the type of $e_g$ orbitals involved and on the direction of the vector linking the Mn ions[23]. In particular, the hopping between a $d_{x^2-y^2}$ orbital and a $d_{3z^2-r^2}$ orbital changes its sign when the hopping is either along the $\hat{x}$ or along the $\hat{y}$ direction. As a result, there are three inequivalent Mn positions along the chain: the corner (C) position, and the x-bridge ($B_x$) and y-bridge ($B_y$) positions. The system opens a gap at the Fermi energy and minimizes the kinetic energy by ordering along the zigzag chain ...C-$B_x$-C-$B_y$-C.., being the occupied $e_g$ orbitals in the form, see Fig.1a,

$$\cdots d_{x^2-y^2} - d_{3x^2-r^2} - d_{x^2-y^2} - d_{3y^2-r^2} - d_{x^2-y^2} \cdots \qquad (1)$$

In this way the combination of AF coupling and directionality in the hopping amplitude produces a spontaneous broken symmetry ground state in the spin and orbital sectors.

In absence of Jahn-Teller coupling, the CE phase can be stabilized with AF interactions between the Mn core spins. The zig-zag chains are uncoupled, and the electronic structure consists of two bands with energy $\pm\frac{4}{3}t\sqrt{2+\cos k}$, where k is the wavevector, and two dispersionless bands at zero energy. At half doping the lowest energy subband is fully occupied. In that case, all the Mn ions have the same electric charge and the gap at the Fermi energy, Fig.2a, is just due to the orbital order and has the value $4/3t$, being $t$ the hopping amplitude between two first neighbour $d_{x^2-y^2}$ orbitals in the x-y plane. For finite $\lambda$, the charge located on a bridge Mn becomes coupled with the distortion along the bridge direction of the oxygen octahedra surrounding the Mn. Therefore Mn ions in corner and bridge positions trap different amount of charge and the system develops a charge density wave with an amplitude that increases with $\lambda$[22-24].

As the gap at half doping is due to orbital ordering, it is convenient to describe this order using a pseudospin picture where $d_{x^2-y^2}$ and $d_{3z^2-r^2}$ $e_g$ states are mapped as pseudospin up and down respectively. With this notation, the orbital ordering along the zigzag chain corresponds to a modulation of the $x$-component of the pseudospin in the form,

$$\tau_x(i) = A\cos\left(\frac{\pi}{2}i + \phi\right) \qquad (2)$$

Here $i$ is the position along the chain and the phase $\phi$ can get the values $\phi = n\pi/2$, being $n$ an integer. This degeneracy reflects the broken symmetry in the orbital sector. Associated to this degeneracy we expect there to exist topological excitations which in this case are solitons. A soliton corresponds to a collective excitation where the phase in Eq.(2) changes slowly from 0 to $\pm\pi/2$. The soliton in the zigzag chain has topological charge $\pm 1/2$[1] and also carries some electrical charge[3]. For a pseudospin modulation $A\cos\left(\frac{\pi}{2}i + \phi\right)$, $\pi/2$ is the Fermi wave vector and is related to the hole concentration, $x$, $\pi/2 = \pi x$, except in the region inside the soliton. Assuming that in a soliton the phase changes slowly, the pseudospin modulation within the soliton can be written as $A\cos(\pi x i + \nabla\phi\, i + \phi_0)$, and the local charge inside the soliton is $\pi(x+\delta)$ with $\delta = \nabla\phi/\pi$. By integrating between 0 and $\pm\pi/2$ we obtain the total charge of the soliton, $Q_s = \pm 1/2$, so that each soliton has an electric charge of half an electron or half a hole.

The equivalence between topological and electric charge suggests that solitons can be the relevant charge excitation in half doped manganites. To verify and quantify this proposal, we compute the energy and properties of the solitons using the microscopic model [22-24] described in Methods. In a given CE chain, we create a soliton by introducing a defect in such a way that the orbital order phase changes in $\pm \pi/2$ when passing through it, see Fig.1. For a $+\pi/2$ change in the orbital phase, the minimum defect needed would correspond to the introduction of an extra corner in the chain,

$$\cdots - C - B_x - C - B_y - \mathbf{C} - C - B_y - C - B_x - \cdots \quad (3)$$

while for a change of $-\pi/2$ the minimum defect is to introduce an extra bridge,

$$\cdots - C - B_x - C - B_y - \mathbf{B_y} - C - B_x - C - B_y - \cdots \quad (4)$$

It is important to note that the defects leading to the appearance of orbital solitons correspond to a rearrangement of the magnetic order, and do not involve any structural or chemical alteration.

Due to the two-dimensional character of the CE phase, each soliton is attached to similar solitons in the neighbouring chains that are coupled AF in order to maintain the magnetic energy unaltered with respect the CE phase, Fig.1. The AF arrangement of the chains precludes the hopping of carriers between chains, and the only inter chain interaction is due to the elastic coupling between Jahn-Teller deformations in neighbouring chains.

In Fig.2 **b,c**, we plot the density of states for the CE phase in the presence of an extra corner or an extra bridge for the case of zero Jahn-Teller coupling for which the chains do not interact. The presence of the defect leads to the appearance of a localized electronic state in the gap. Half of the spectral weight of the gap state arises from the valence band and the other half from the conduction band. Therefore, by occupying the gap state we add half electron per defect to the system. On the contrary, if the Fermi energy is located between the valence band and the gap state we would be adding half hole per defect to the system. In Fig.3, we plot the spatial variation of the orbital phase $\phi$ for a system containing an extra half hole and the defect described by Eq.(4). The phase can be fitted by $\phi(i) = \frac{\pi}{2} \tanh\left(-\frac{i}{i_0}\right)$, with $i_0 \approx 2.25$. This indicates that the system screens the extra bridge by creating a rather smooth soliton. Similar results are obtained when the defect is created by an extra corner. We therefore conclude that in order to screen the defects described by Eq. (3) and Eq.(4) the system creates positive and negative solitons in the orbital order. The lack of symmetry around $x=1/2$ makes the energy of the positive and negative solitons to be different, in such a way that extra electrons prefer to create positive solitons while extra holes rather create negative solitons.

In the case of finite $\lambda$, there is an elastic interaction between neighbouring chains, and the soliton energy increases with the Jahn-Teller coupling. The elastic coupling occurs because each oxygen involved in the octahedron distortion around a Mn ion is also part of the octahedron surrounding another Mn ion. In Fig.4 we plot, for the half doped CE phase, the energy corresponding to the creation of two negative charged solitons compared to the energy of adding to the system a hole maintaining the CE symmetry. In the latter case the energy is basically the gap energy of the CE phase. For small values of $\lambda$, the interchain soliton interaction is rather weak and the added charge creates charged solitons. For larger values of $\lambda$ the elastic repulsion between solitons in neighbouring chains increases and eventually, for large values of $\lambda$, the system prefers to accommodate the extra charge in the CE phase, instead of creating topological excitations. From Fig.4 we conclude that there is a wide range of values of $\lambda$, for which the relevant charge excitations in the half doped CE phase are solitons.

A finite amount of charge added to the CE phase creates a finite density of solitons at each zig-zag chain and the system forms an incommensurate solitonic phase. This solitonic phase is formed by and array of half charged solitons separated by $1/2|x-0.5|$ ions along the CE chains and has an orbital modulation of wave vector $\mathbf{q}=(1-x)\mathbf{a}^*$. In order to prove that the solitonic phase is the ground state near $x=1/2$, we compare its energy with the energies of the ferromagnetic metallic phase and the CE commensurate phase. In Fig.5 we plot the phase diagram $x$-$\lambda$ for $J_{AF}=0.1t$. In the range $0.45<x<0.55$ the solitons along the chains practically do not interact and can be considered as isolated defects. For values of $x$ outside this range the ground state can not be described by isolated solitons and more sophisticated phases as bistripes or diagonal phases appear[24,26,27]. The phase diagram in Fig.5 clearly shows the asymmetry around half doping. At weak Jahn-Teller coupling the delocalising kinetic energy is the most important interaction and the ground state is ferromagnetic and metallic. Only for $\lambda$ larger than a critical value $\lambda_c$, an ordered phase occurs. The value of $\lambda_c$ decreases when the hole concentration increases, as the relative strength of the kinetic energy is reduced. Aside from the commensurate phase at $x=1/2$, for moderate values of $\lambda$ the ordered phase is the incommensurate solitonic phase. For larger values of the Jahn-Teller coupling, the phase diagram shows again an opposite behaviour to the addition of electrons or holes. Due to the lack of symmetry in the CE density of states versus doping, extra electrons prefer to accommodate in the commensurate phase whereas extra holes create solitons forming a incommensurate phase. For larger or smaller values of $J_{AF}$, we obtain similar phase diagrams, with smaller or larger values of $\lambda_c$ respectively.

The value of $\lambda/t$ determines the charge modulation in the manganite. In our model, values of $\lambda/t$ of 1, 1.2, 1.6 and 1.8 imply a charge disproportion between Mn ions of 0.14, 0.23, 0.46 and 0.58 electrons respectively. Recent X-ray resonant scattering experiments indicate[28,29] that charge modulation between the Mn ions is only a small fraction of the average charge density, the picture being rather different to a distribution of pure $Mn^{+3}$ and $Mn^{+4}$. Therefore, we estimate that the Jahn-Teller coupling should be smaller than $1.8t$, and its precise value should depend on the bandwidth of the particular manganites, being $\lambda/t$ small for $LaSrMnO_3$ and larger for $LaCaMnO_3$ and $PrCaMnO_3$. We propose that for $LaSrMnO_3$ the value of the Jahn-Teller coupling is smaller than $\lambda_c$, thus explaining that doping either with electrons or holes results in a metallic ground state[15]. On the contrary, for $LaCaMnO_3$ and $PrCaMnO_3$, the value of $\lambda$ is high enough so that the $x=0.5$ doping line separates commensurate from incommensurate phases[8,11].

Two experimental consequences derive from the existence of topological defects in the CE phase. First, contrary to the Sine-Gordon like functional predictions, there is not a critical lock-in density of holes for the occurrence of a commensurate –incommensurate transition as any amount of extra holes in the system forms an incommensurate solitonic phase. And second, the states that appear in the middle of the gap in the incommensurate solitonic phase, Fig.3, should modify strongly the optical conductivity, and at energies smaller than the CE energy gap, absorption peaks would emerge when moving from half filling.

The existence of charged topological defects associated with orbital order explains the asymmetry around half doping without requiring external sources of disorder or strain[13-14]. Models based in Ginzburg-Landau phenomenology[30] explain inhomogeneities in manganites as a result of coupling between order parameters. In that formalism the electronic part of the free energy is eliminated by treating only with locally uniform order parameters. However, the appearance of solitons with doping modifies drastically the electronic spectrum from the one for a uniform case and the electronic part cannot be eliminated at the outset if charged incommensurate solitonic phases want to be considered in a Ginzburg-Landau formalism.

In closing, the broken orbital symmetry ground state occurring in half doped manganites supports the existence of charged solitons in the orbital sector. For a wide range of Jahn-Teller and antiferromagnetic couplings, these excitations could be the relevant charged excitations in the system in such a way that added charges form an incommensurate solitonic phase. For moderate values of the Jahn-Teller coupling we obtain a strong asymmetry between adding electrons that keeps the system commensurate or adding holes, that produces an incommensurate phase. Orbital solitons may be a general phenomena occurring at other doping in manganites and in other strongly correlated systems[31].

METHODS

The microscopic model we use to describe the manganites, is a two orbital double exchange model coupled with Jahn-Teller phonons[22-24]. The Hamiltonian of the system is the sum of the following terms:

i) The kinetic energy term that describes the hopping between Mn ions that we consider finite for first neighbours. The hopping amplitude, $t_{\alpha,\alpha'}^{u}$, depends on the type of orbitals involved and on the direction, $u$, between sites; $t_{a,a}^{x(y)} = \pm\sqrt{3} t_{a,b}^{x(y)} = \pm\sqrt{3} t_{b,a}^{x(y)} = t$ and $t_{b,b}^{z} = \frac{4}{3}t$, with $t_{a,a}^{z} = t_{a,b}^{z} = t_{b,a}^{z} = 0$, where a and b denote the orbitals $d_{x^2-y^2}$ and $d_{3z^2-r^2}$ respectively. In the large Hund's coupling limit, the hopping amplitude between two Mn ions is weighted [20] by the double exchange spin reduction factor $\sqrt{\frac{1+\cos^2\theta}{2}}$, where $\theta$ is the angle formed by the Mn core spins,

ii) An electron phonon coupling between the $e_g$ electrons and the three active $MnO_6$ octahedra distortions: the breathing mode $Q_1$ and the Jahn Teller modes $Q_2$ and $Q_3$ that have symmetry $x^2$-$y^2$ and $3z^2$-$r^2$ respectively.

iii) The elastic energy of the octahedra distortions, $\frac{1}{2}(Q_1^2 + Q_2^2 + Q_3^2)$.

We also include the antiferromagnetic exchange coupling among the $t_{2g}$ Mn spins, $J_{AF} \sum_{<i,j>} \vec{S}_i \vec{S}_j$, and a Hubbard term of magnitude $U'$, that accounts for the interorbital Coulomb interaction and penalizes the occupancy of two orbitals at the same site. In the mean field approximation, it has been proved [22], that on site interorbital Coulomb interaction plays the same role as the Jahn-Teller coupling, and it is not necessary to include U' explicitly in the calculations. On the other hand, in the limit of infinite Hund's coupling, two electrons can not be located in the same orbital at the same site, and the double occupancy is suppressed, being unnecessary the inclusion of the onsite Hubbard term.

In the perovskite structure the oxygen atoms are shared by neighbouring $MnO_6$ octahedra, and the Q's distortions are not independent, being cery important cooperative effects. We consider this collective effect by taking the position of the oxygen atoms as the independent variables of the Jahn-Teller distortions.

CALCULATIONS

For a particular number of carriers, a given set of values $\lambda$ and $J_{AF}$ and a texture of core spins, we solve self-consistently the mean field version of the Hamiltonian and obtain energy, local charges and orbital orientations. We have checked that the spatial dimensions of the systems studied are large enough to preclude finite size effects.

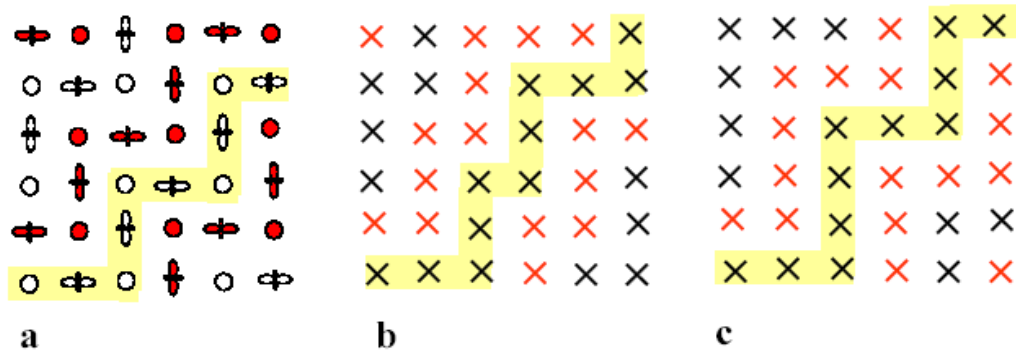

Fig.1. Magnetic and orbital order. **a**. Orbital and spin order of half doped manganites in the *x-y* plane. Elongated orbitals along the *x* and *y* directions represent $d_{3x^2-r^2}$ and $d_{3y^2-r^2}$ orbitals respectively. Circles represent $d_{x^2-y^2}$ orbitals. Open and solid symbols indicate up and down Mn core spins. Shaded region emphasizes a zigzag chain. **b**. Spin texture of a phase formed by parallel zigzag chains each of them with a positive soliton. Shaded region emphasizes the chain indicated by Eq.3. **c**. Spin texture of a phase formed by parallel zigzag chains each of them with a negative soliton. Shaded region emphasizes the chain indicated by Eq.4. In **b** and **c** red and black crosses indicates up and down Mn core spins.

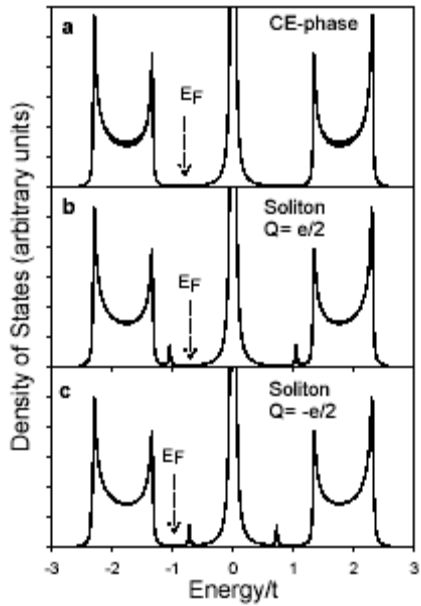

Fig.2. Density of states in the case of zero Jahn Teller coupling. **a**. In the perfect CE phase the system develops a gap that separates the occupied from the empty states. **b** and **c**. Density of states of the CE phase in presence of a density of positive and negative solitons respectively. The calculations have been done for a periodic distribution of non-interacting widely separated defects. The arrows indicates the position of the Fermi energy in the CE phase at *x*=1/2 (**a**), in presence of half extra electron per soliton (**b**) and in presence of half extra hole per soliton (**c**).

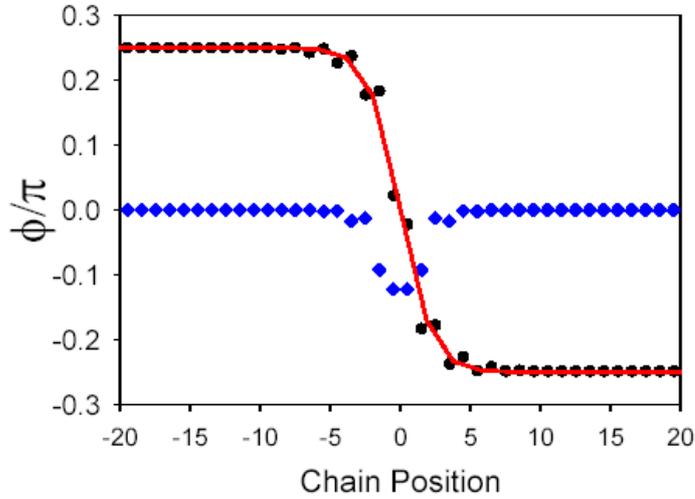

Fig.3. Shape of the orbital soliton as function of the position. The dots represent the orbital phase as obtained by minimizing the kinetic energy of the CE chain containing a bridge defect and an extra half hole. The index $i$ indicates the position along the chain. The soliton is centered at $i=0$. The continuous line corresponds to the fitting $\phi(i) = \dfrac{\pi}{2}\tanh\left(-\dfrac{i}{i_0}\right)$ with $i_0 \approx 2.25$. The density of electric charge associated with the presence of the soliton is represented by diamonds. There is a deficit of half electron associated with the presence of the soliton.

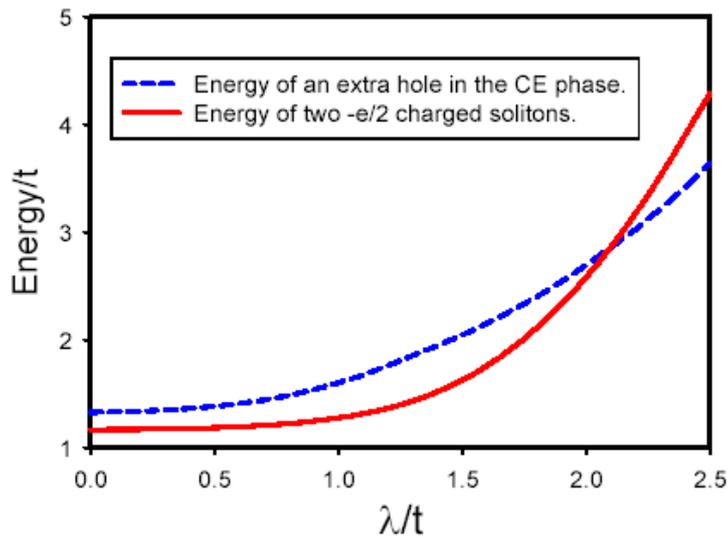

Fig.4. Energy of charge excitations of the half doped CE phase as function of the Jahn-Teller coupling $\lambda$. The energy cost of adding an hole in the CE phase is roughly the energy gap and increases with the electron lattice coupling. The energy cost of creating the topological excitations increases stronger with $\lambda$ as a result of the interchain elastic interaction.

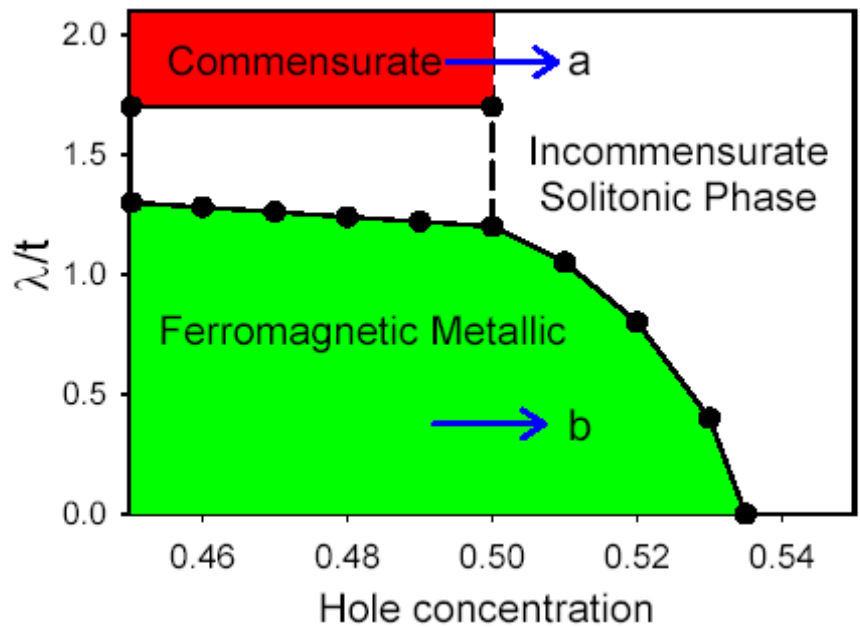

Fig.5. Phase diagram, *x*-λ, for $J_{AF}$=0.1*t* that results from solving the two orbital double exchange hamiltonian. Red region represents the CE commensurate phase. The white area corresponds to the incommensurate phase, whereas the green region is the metallic ferromagnetic phase. Arrow **a** indicates the commensurate incommensurate transition in materials with a moderate charge modulation as $PrCaMnO_3$ and $LaCaMnO_3$. In materials with a small Jahn-Teller coupling, as $LaSrMnO_3$, the ground state near half doping is metallic and ferromagnetic, arrow **b**.

**Acknowledgements**. The authors thank M.J.Calderón, P.López-Sancho, N.D.Mathur A. Ruiz and C.Tejedor for helpful discussions. LB thanks Cambridge University for the hospitality during the realization of this work. Financial support is acknowledged from Grants No MAT2002-04429-C03-01 (MCyT, Spain) and Fundación Ramón Areces. The work at Cambridge was partially supported by the UK EPSRC.